\documentclass[11pt]{article}
\usepackage[russian]{babel}
\usepackage{amssymb,amsmath,amsthm}
\newcommand{\bc}{\begin{center}}
\newcommand{\ec}{\end{center}}
\newcommand{\be}{\begin{equation}}
\newcommand{\ee}{\end{equation}}
\newcommand{\bea}{\begin{eqnarray}}
\newcommand{\eea}{\end{eqnarray}}
\newcommand{\ba}{\begin{array}}
\newcommand{\ea}{\end{array}}
\newcommand{\edc}{\end{document}}

\def\s{\sigma}

\begin{document}
УДК 517.98
\begin{center}
\textbf{\Large {Новые периодические меры Гиббса для модели Поттса с $q-$состояниями на дереве Кэли}}\\
\end{center}

\begin{center}
\textbf{\Large {Р.М.Хакимов}}\\
\end{center}

\begin{center}
Институт математики, ул. Дурмон йули, 29,
Ташкент, 100125, Узбекистан.\\
E-mail: rustam-7102@rambler.ru
\end{center}

В данной статье изучается модель Поттса с $q-$состояниями на
дереве Кэли порядка $k$ и показано существование периодических (не
трансляционно-инвариантных) мер Гиббса при некоторых условиях на
параметры этой модели. Кроме того, указана нижняя граница
количества существующих периодических
мер Гиббса.\\

\textbf{Ключевые слова}: дерево Кэли, конфигурация, модель Поттса,
мера Гиббса, периодические меры, трансляционно-инвариантные меры.

\section{Введение}\

Основной задачей гамильтониана является описание всех отвечающих
ему предельных мер Гиббса. Определение меры Гиббса и понятия,
связанных с этой теорией, вводятся стандартным образом (см.
например \cite {6}-\cite {Si}). Для модели Изинга на дереве Кэли
эта задача изучена достаточно полно. Так, например, в работе
\cite{Bl} построено несчетное множество крайних гиббсовских мер, а
в работе \cite{Bl1} найдено необходимое и достаточное условие
крайности неупорядоченной фазы модели Изинга на дереве Кэли.

Модель Поттса является обобщением модели Изинга. Несмотря на
многочисленность работ, посвященных модели Поттса на дереве Кэли,
она изучена мало по сравнению с моделью Изинга. Так, например, в
работе \cite{Ga8} изучена ферромагнитная модель Поттса с тремя
состояниями на дереве Кэли второго порядка и показано
существование критической температуры $T_c$ такой, что при $T<T_c$
существуют три трансляционно-инвариантных и несчетное число не
трансляционно-инвариантных мер Гиббса. В работе \cite {GN}
обобщены результаты работы \cite{Ga8} для модели Поттса с конечным
числом состояний на дереве Кэли произвольного (конечного) порядка.

В работе \cite{1} доказано, что на дереве Кэли
трансляционно-инвариантная мера Гиббса антиферромагнитной модели
Поттса с внешним полем единственна. Работа \cite{Ga13} посвящена
модели Поттса со счетным числом состояний и c ненулевым внешним
полем на дереве Кэли. Доказано, что эта модель имеет единственную
трансляционно-инвариантную меру Гиббса.

Для ознакомления с другими свойствами модели Поттса на дереве Кэли
смотрите, например, монографию \cite{Rb}. В работе \cite{RK}
показано, что модель Поттса (с внешним полем $\alpha\in R$) имеет
только периодические меры Гиббса с периодом два; рассмотрен случай
$\alpha=0$, и на некоторых инвариантах доказано, что все
периодические меры Гиббса являются трансляционно-инвариантными;
найдены условия, при которых модель Поттса с ненулевым внешним
полем имеет периодические (не трансляционно-инвариантные) меры
Гиббса. В работе \cite {KRK} дано полное описание
трансляционно-инвариантных мер Гиббса для ферромагнитной модели
Поттса с $q-$состояниями и показано, что их  количество не больше,
чем $2^q-1.$ В работе \cite{Khakimov1} для модели Поттса с тремя
состояниями и с нулевым внешним полем на дереве Кэли порядка $k=3$
и $k=4$ при некоторых условиях на параметры на некоторых
инвариантах показано существование периодических (не
трансляционно-инвариантных) мер Гиббса с периодом два, а в работе
\cite{Khakimov2} для модели Поттса с тремя состояниями и с нулевым
внешним полем на дереве Кэли порядка $k\geq3$ показано
существование критической температуры $T_c$ такой, что при $T<T_c$
существуют не менее двух периодических (не
трансляционно-инвариантных) мер Гиббса.

В этой работе мы покажем существование периодических (не
трансляционно-инвариантных) мер Гиббса на некоторых инвариантах
при некоторых условиях на параметры модели Поттса с
$q-$состояниями и с нулевым внешним полем на дереве Кэли порядка
$k$ и укажем такое число, что количество периодических (не
трансляционно-инвариантных) мер Гиббса не меньше этого числа.

\section{Определения и известные факты}\

Дерево Кэли $\Im^k$ порядка $ k\geq 1 $ - бесконечное дерево, т.е.
граф без циклов, из каждой вершины которого выходит ровно $k+1$
ребер. Пусть $\Im^k=(V,L,i)$, где $V$ есть множество вершин
$\Im^k$, $L$ - его множество ребер и $i$ - функция инцидентности,
сопоставляющая каждому ребру $l\in L$ его концевые точки $x, y \in
V$. Если $i (l) = \{ x, y \} $, то $x$ и $y$ называются  {\it
ближайшими соседями вершины} и обозначаются через $l = \langle x,
y \rangle $. Расстояние $d(x,y), x, y \in V$ на дереве Кэли
определяется формулой
$$
d (x, y) = \min \ \{d | \exists x=x_0, x_1,\dots, x _ {d-1},
x_d=y\in V \ \ \mbox {такой, что} \ \ \langle x_0,
x_1\rangle,\dots, \langle x _ {d-1}, x_d\rangle\} .$$

Для фиксированого $x^0\in V$ обозначим $ W_n = \ \{x\in V\ \ | \ \
d (x, x^0) =n \}, $
\begin{equation}\label{p*}
 V_n = \ \{x\in V\ \ | \ \ d (x, x^0) \leq n \},\ \ L_n = \ \{l =
\langle x, y\rangle \in L \ \ | \ \ x, y \in V_n \}.
\end{equation}

Известно, что существует взаимно-однозначное соответствие между
множеством $V$ вершин дерева Кэли порядка $k\geq 1 $ и группой $G
_{k},$ являющейся свободным произведением $k+1$ циклических групп
второго порядка с образующими $a_1, a_2,\dots, a_{k+1} $,
соответсвенно.\

Мы рассмотрим модель, где спиновые переменные принимают значения
из множества $\Phi = \ \{1, 2,\dots, q \},$ $ q\geq 2 $ и
расположены на вершинах дерева. Тогда \emph{ конфигурация} $\s$ на
$V$ определяется как функция $x\in V\to\s (x) \in\Phi$; множество
всех конфигураций совпадает с $\Omega =\Phi ^ {V} $.

Гамильтониан модели Поттса определяется как
$$H(\sigma)=-J\sum_{\langle x,y\rangle\in L}
\delta_{\sigma(x)\sigma(y)},\eqno(2)$$ где $J\in R$, $\langle
x,y\rangle-$ ближайшие соседи и $\delta_{ij}$-символ Кронекера:
$$\delta_{ij}=\left\{\begin{array}{ll}
0, \ \ \mbox{если} \ \ i\ne j\\[2mm]
1, \ \ \mbox{если} \ \ i= j.
\end{array}\right.
$$
Определим конечномерное распределение вероятностной меры $\mu$ в
обьеме $V_n$ как $$\mu_n(\sigma_n)=Z_n^{-1}\exp\left\{-\beta
H_n(\sigma_n)+\sum_{x\in W_n}h_{\sigma(x),x}\right\},\eqno(3)$$
где $\beta=1/T$, $T>0$-температура,  $Z_n^{-1}$-нормирующий
множитель и $\{h_x=(h_{1,x},\dots, h_{q,x})\in R^q, x\in V\}$-
совокупность векторов и
$$H_n(\sigma_n)=-J\sum_{\langle x,y\rangle\in L_n}
\delta_{\sigma(x)\sigma(y)}.$$

Говорят, что вероятностное распределение (3) согласованное, если
для всех $n\geq 1$ и $\sigma_{n-1}\in \Phi^{V_{n-1}}$:
$$\sum_{\omega_n\in \Phi^{W_n}}\mu_n(\sigma_{n-1}\vee
\omega_n)=\mu_{n-1}(\sigma_{n-1}).\eqno(4)$$

Здесь $\sigma_{n-1}\vee \omega_n$  есть объединение конфигураций.
В этом случае, существует единственная мера $\mu$ на $\Phi^V$
такая, что для всех $n$ и $\sigma_n\in \Phi^{V_n}$
$$\mu(\{\sigma|_{V_n}=\sigma_n\})=\mu_n(\sigma_n).$$
Такая мера называется расщепленной гиббсовской мерой,
соответствующей гамильтониану (2) и векторнозначной функции $h_x,
x\in V$.\

Следующее утверждение описывает условие на $h_x$, обеспечивающее
согласованность $\mu_n(\sigma_n)$.

\textbf{Теорема 1}.\cite{1} \textit{Вероятностное распределение
$\mu_n(\sigma_n)$, $n=1,2,\ldots$ в (3) является согласованной
тогда и только тогда}, \textit{когда для любого} $x\in V$
\textit{имеет место следующее
$$h_x=\sum_{y\in
S(x)}F(h_y,\theta),\eqno(5)$$ где $F: h=(h_1,
\dots,h_{q-1})\in R^{q-1}\to
F(h,\theta)=(F_1,\dots,F_{q-1})\in R^{q-1}$ определяется
как:
$$F_i=\ln\left({(\theta-1)e^{h_i}+\sum_{j=1}^{q-1}e^{h_j}+1\over
\theta+ \sum_{j=1}^{q-1}e^{h_j}}\right),$$ и
$\theta=\exp(J\beta)$, $S(x)-$ множество прямых потомков точки
$x$.}\\

Пусть $\widehat{G}_k-$ подгруппа группы $G_k$.

\textbf{Определение 1}. Совокупность векторов $h=\{h_x,\, x\in
G_k\}$ называется $ \widehat{G}_k$-периодической, если
$h_{yx}=h_x$ для $\forall x\in G_k, y\in\widehat{G}_k.$

$G_k-$ периодические совокупности называются
трансляционно-инвариантными.

\textbf{Определение 2}. Мера $\mu$ называется
$\widehat{G}_k$-периодической, если она соответствует
$\widehat{G}_k$-периодической совокупности векторов $h$.

Следуюшая теорема характеризует периодические меры Гиббса.

\textbf{Теорема 2.}\cite{RK} \textit{Пусть $K-$ нормальный делитель
конечного индекса в $G_k.$ Тогда для модели Поттса все $K-$
периодические меры Гиббса являются либо $G_k^{(2)}-$
периодическими, либо трансляционно-инвариантными.}

\section{Периодические меры Гиббса}\

Рассмотрим случай $q\geq 3$, т.е. $\sigma:V\rightarrow\Phi=
\{1,2,3,...,q\}$. В силу Теоремы 2 имеются только
$G^{(2)}_k$-периодические меры Гиббса, которые соответствуют
совокупности векторов $h=\{h_x\in R^{q-1}: \, x\in G_k\}$ вида
$$h_x=\left\{%
\begin{array}{ll}
    h, \ \ \ $ если $ |x|-\mbox{четно} $,$ \\
    l, \ \ \ $ если $ |x|-\mbox{нечетно} $.$ \\
\end{array}%
\right. $$
 Здесь $h=(h_1,h_2,...,h_{q-1}),$ $l=(l_1,l_2,...,l_{q-1}).$
Тогда в силу (5) имеем:
$$
\left\{%
\begin{array}{ll}
    h_{i}=k\ln{(\theta-1)\exp(l_i) + \sum_{j=1}^{q-1}exp({l_j})+1\over \sum_{j=1}^{q-1}exp({l_j})+\theta},\\[3 mm]
    l_{i}=k\ln{(\theta-1)\exp(h_i) + \sum_{j=1}^{q-1}exp({h_j})+1\over \sum_{j=1}^{q-1}exp({h_j})+\theta},  \\
\end{array}%
i=\overline{1,q-1}. \right.$$\

Введем следующие обозначения: $\exp(h_i)=x_i,\ \exp(l_i)=y_i.$
Тогда последнюю систему уравнений при $i=\overline{1,q-1}$ можно
переписать:
$$
\left\{%
\begin{array}{ll}
    x_{i}=\left({(\theta-1)y_i + \sum_{j=1}^{q-1}y_j+1\over \sum_{j=1}^{q-1}y_j+\theta}\right)^k, \\[3 mm]
    y_{i}=\left({(\theta-1)x_i + \sum_{j=1}^{q-1}x_j+1\over \sum_{j=1}^{q-1}x_j+\theta}\right)^k.\\
    \end{array}%
\right.\eqno(6)$$\

\textbf{Замечание 1.} 1) При $q=2$ модель Поттса совпадает с
моделью Изинга, которая была изучена в работе \cite{1}.

2) В случае $k=2, \ q=3$ и $J<0$ было доказано, что на
инвариантном множестве $I=\{(x_1,x_2,y_1,y_2)\in R^4: x_1=x_2, \
y_1=y_2\}$ все $G_k^{(2)}-$периодические меры Гиббса являются
трансляционно-инвариантными (см. \cite{RK}).

3) В случае $k\geq 1, \ q=3$ и $J>0$ было доказано, что все
$G_k^{(2)}-$периодические меры Гиббса являются
трансляционно-инвариантными (см. \cite{RK}).

При $q\geq 3,\ 0<\theta <1, \ k\geq 3$ рассмотрим множества
$$I_m=\{z=(u,v)\in R^{q-1}\times R^{q-1}: x_i=x, \ y_i=y,
i=\overline{1,m}, \ x_i=y_i=1, i=\overline{m+1,q-1}\};$$ т.е.
$u=(\underbrace{x,x,...,x}_m,1,1,...,1), \
v=(\underbrace{y,y,...,y}_m,1,1,...,1)$ и
$$I_m^{'}=\{z=(u,v)\in R^{q-1}\times R^{q-1}: x_i=x, \ i=\overline{1,m},\ x_i=1, \ i=\overline{m+1,q-1-m},$$
$$x_i=y, i=\overline{q-m,q-1}, y_i=y, \ i=\overline{1,m},\ y_i=1, \ i=\overline{m+1,q-1-m}, y_i=x,i=\overline{q-m,q-1} \},$$ т.е.
$u=(\underbrace{x,x,...,x}_m,1,1,...,1,\underbrace{y,y,...,y}_m),
\
v=(\underbrace{y,y,...,y}_m,1,1,...,1,\underbrace{x,x,...,x}_m).$
Здесь $2m\leq q-1.$\

Рассмотрим отображение $W:R^{q-1}\times R^{q-1} \rightarrow
R^{q-1}\times R^{q-1},$ определенное следующим образом:
$$
\left\{%
\begin{array}{ll}
    x_{i}^{'}=\left({(\theta-1)y_i + \sum_{j=1}^{q-1}y_j+1\over \sum_{j=1}^{q-1}y_j+\theta}\right)^k \\[3 mm]
    y_{i}^{'}=\left({(\theta-1)x_i + \sum_{j=1}^{q-1}x_j+1\over \sum_{j=1}^{q-1}x_j+\theta}\right)^k.\\
    \end{array}%
\right.$$ Заметим, что (6) есть уравнение $z=W(z).$ Чтобы решить
систему уравнений (6), надо найти неподвижные точки отображения
$z^{'}=W(z),$ где $z=(u,v), \ z^{'}=(u^{'},v^{'}).$\

\textbf{Лемма.} Множества $I_m$ и $I_m^{'}$ являются инвариантными
множествами относительно отображения $W$ .

Доказательство аналогично доказательству леммы 2 из \cite {RK}.\\

\textbf{Случай $I_m$.} В этом случае система уравнений (6) имеет
следующий вид:
$$
\left\{%
\begin{array}{ll}
    x=\left({\theta y +(m-1)y+(q-m)\over \theta +m y+(q-m-1)}\right)^k \\[3 mm]
    y=\left({\theta x +(m-1)x+(q-m)\over \theta +m x+(q-m-1)}\right)^k\\
    \end{array}%
\right.\eqno(7)$$\ или
$$
\left\{%
\begin{array}{ll}
    x=f^k(y) \\
    y=f^k(x), \\
    \end{array}%
\right. \texttt{где}\ f(x)={\theta x +(m-1)x+(q-m)\over \theta +m
x+(q-m-1)}, \eqno(8)$$ и $f^k(x)$ есть $k$-ая степень функции
$f(x).$

\textbf{Замечание 2.} Пусть $\pi\in S_{q-1}$ перестановка.
Определим действие $\pi$ на вектор $x=(x_1,x_2,...,x_{q-1})$ как
$\pi (x)=(x_{\pi(1)},x_{\pi(2)},...,x_{\pi(q-1)})$. Тогда
$\pi(A)=\{(\pi x,\pi y): (x,y)\in A \}$, где $A=I_m$ или
$I_m^{'}$, также является инвариантным множеством относительно
$W$, но соответствующая система уравнений в случае $I_m$ совпадает
с (7), а в случае $I_m^{'}-$ с (9), поэтому не нарушая общности,
можно рассмотреть $I_m$ и $I_m^{'}.$

\textbf{Утверждение 1.} \textit{Пусть} $k\geq 3$,\ $3\leq
q<k+1$,\, $\theta_{cr}=\frac{k-q+1}{k+1}<1$. \textit{Тогда система
уравнений (6) на инвариантном множестве $I_m$ при
$0<\theta<\theta_{cr}$ имеет не менее трех решений, при
$\theta=\theta_{cr}$ имеет не менее одного решения и при $\theta >
\theta_{cr}$ имеет только одно решение}.\

\textbf{Доказательство.} Из (8) получим
$$x=g(x)=f^k(f^k(x)).$$
Мы имеем
$$ f'(x)={(\theta-1)(\theta+q-1)\over (\theta+my+q-m-1)^2};$$
$$g'(x)=k^2f^{k-1}(f^k(x))f'(f^k(x))f^{k-1}(x)f'(x);$$
Из этих формул следует, что функция $f(x)$ убывает при
$0<\theta<1$ и $f(x)=x$ имеет единственное решение $x=1$, причем
$f'(1)={\theta-1\over \theta+q-1}$. Заметим, что $g(x)$ возрастает
и $x=1$ является решением $g(x)=x$. Это решение не единственно,
если $g'(1)=k^2(f'(1))^2=\left(k{\theta-1\over
\theta+q-1}\right)^2>1$, т.к. в этом случае график функции $g$ при
$x>1$ будет лежать выше биссектрисы и функция $g$ ограничена.
Таким образом, критическое значение для $\theta$ получается из
уравнения $\left(k{\theta-1\over \theta+q-1}\right)^2=1$, которое
при $\theta<1$ дает $\theta_{cr}=\frac{k-q+1}{k+1}$. Из этих
рассуждений следует, что при $0<\theta<\theta_{cr}$ уравнение
$g(x)=x$ имеет не менее трех решений $x_0^*<x_1^*=1<x_2^*,$ т.е.
уравнение $g(x)=x$ имеет не менее двух корней, отличных от корней
уравнения $f(x)=x$. При $\theta=\theta_{cr}$ график функции $g$
касается биссектрисы в точке $x=1$. Это значит, что уравнение
$g(x)=x$ при этом условии имеет не менее одного решения. Кроме
того, ясно, что при $\theta>\theta_{cr}$ уравнение $g(x)=x$ имеет
единственное решение $x_1^*=1$, которое является решением
уравнения $f(x)=x$. Утверждение доказано.\\

\textbf{Случай $I_m^{'}$.} Рассмотрим множество $I_m^{'}$. Система
уравнений (6) после обозначений $\exp(h_i)=x_i,\ \exp(l_i)=y_i$ на
этом множестве имеет вид

$$
\left\{%
\begin{array}{ll}
    x=\left({(\theta-1) y +my+(q-2m-1)+mx+1\over \theta +mx+m y+(q-2m-1)}\right)^k \\[3 mm]
    y=\left({(\theta-1) x +mx+(q-2m-1)+my+1\over \theta +m x+my+(q-2m-1)}\right)^k\\
    \end{array}%
\right.\eqno(9)$$

\textbf{Замечание 3.} 1) При $m=0$ получим $u=(1,1,...,1), \
v=(1,1,...,1)$, которые соответствуют трансляционно-инвариантным
мерам Гиббса, поэтому мы рассмотрим случай $m\geq1$.

2) В случае $k=2, \ q=3, \ m=1$ на $I_m^{'}$ было доказано, что
все $G_k^{(2)}-$периодические меры Гиббса являются
трансляционно-инвариантными (см. \cite{RK}).

Заменив в последней системе уравнений $\sqrt[k]{x}=z, \
\sqrt[k]{y}=t$, получим
$$
\left\{%
\begin{array}{ll}
    z={(\theta +m-1) t^k +mz^k+q-2m\over \theta +mz^k+m t^k+q-2m-1} \\[3 mm]
    t={(\theta +m-1) z^k +mt^k+q-2m\over \theta +m z^k+mt^k+q-2m-1}\\
    \end{array}%
\right.\eqno(10)$$ Из первого уравнения (10) найдем $t^k, \ t$
$$t^k={mz^{k+1}-mz^k+(\theta +q-2m-1)z-q+2m\over \theta +m-1-mz};$$
$$t=\left({mz^{k+1}-mz^k+(\theta +q-2m-1)z-q+2m\over \theta +m-1-mz}\right)^{{1\over k}}$$
и подставим во второе уравнение из (10). В результате получим
$$f(z)=[(\theta+2m-1)z^k-mz^{k+1}+mz+q-2m]^k(\theta+m-1-mz)-$$
$$-(mz^k+q-m-1+\theta)^k[mz^{k+1}-mz^k+(\theta+q-2m-1)z-q+2m]=0.\eqno(11)$$
Рассмотрим функцию $f(z).$ Заметим, что
$f(0)=(q-2m)^k(\theta+m-1)+(q-2m)(\theta+q-m-1)>0$ при $2m<q,$ но
это так потому, что число $q-1-2m>0.$ Кроме того, $f(1)=0$ и
$f(z)\rightarrow -\infty$ при $z \rightarrow +\infty.$ Отсюда
ясно, что если $f'(1)>0$, то уравнение (11) имеет по меньшей мере
три решения. Поэтому рассмотрим
$$f'(1)=(k^2-1)s^2-2qs-q^2=(k^2-1)\left(s-{q\over k-1}\right)\left(s+{q\over k+1}\right)>0,$$ где $s=\theta-1<0.$
Следовательно, $f'(1)>0,$ если $s+{q\over k+1}<0,$ то есть при
$0<\theta<1-{q\over k+1}=\theta_{cr}.$

Таким образом, справедливо утверждение, подобное утверждению 1.

\textbf{Утверждение 2.} \textit{Пусть} $k\geq 3$,\ $3\leq
q<k+1$,\, $\theta_{cr}=\frac{k-q+1}{k+1}<1$. \textit{Тогда система
уравнений (6) на инвариантном множестве} $I_m^{'}$

{1. При $0<\theta<\theta_{cr}$ имеет не менее трех решений};

{2. При $\theta=\theta_{cr}$ имеет не менее одного решения};

{3. При $\theta>\theta_{cr}$ имеет только одно решение}. \

\textbf{Замечание 4.} Ясно, что в утверждениях 1 и 2 в случаях
$0<\theta<\theta_{cr}$ и при $\theta=\theta_{cr}$ одна из
соответствующих мер решению системы (6) ($x_1^*=1$) является
трансляционно-инвариантной (если при $\theta=\theta_{cr}$
существуют меры, отличные от меры, соответствующей решению
$x_1^*=1$), остальные же $G_k^{(2)}-$ периодическими (не
трансляционно-инвариантными), а в случае $\theta>\theta_{cr}$
мера, соответствующая единственному решению $x_1^*=1$, является
трансляционно-инвариантной.\

Далее, подобно работе \cite {KRK} (стр.6), легко показать, что при
$0<\theta<\theta_{cr}$ на каждом $I_m$ и $I_m^{'}$, где
$m=1,2,...,q,$ количество $G_k^{(2)}-$ периодических (не
трансляционно-инвариантных) мер Гиббса не меньше, чем $2C_{q}^m$ и
$2C_{q}^{m}C_{q-m}^{m}$, соответственно. Следовательно, это
количество на $\bigcup_{m=1}^{q} I_m$ и $\bigcup_{m=1}^{q}
I_m^{'}$ не меньше, чем
$$2 \sum_{m=1}^{q}C_{q}^{m}=2(C_{q}^1+C_{q}^2+...+C_{q}^{q})=2^{q+1}-2,$$
и
$$2 \sum_{m=1}^{[q/2]}C_{q}^{m}C_{q-m}^{m},$$
соответственно.

Таким образом, справедлива следующая

\textbf{Теорема 3.} {Для модели Поттса при $k\geq 3$,\ $3\leq
q<k+1$ и $0<\theta<\theta_{cr}$ существуют не менее
$$2\cdot \left(2^{q}-1+ \sum_{m=1}^{[q/2]}C_{q}^{m}C_{q-m}^{m}\right)$$
$G_k^{(2)}-$ периодических (не трансляционно-инвариантных) мер
Гиббса}.\

\textbf{Замечание 5.} Количество и описание всех
трансляционно-инвариантных мер Гиббса для модели Поттса были
изучены в работе \cite {KRK}.\

\textbf{Благодарность.} Автор выражает глубокую признательность
профессору У. А. Розикову за постановку задачи и полезные советы
по работе.

\end{document}